\documentclass[onecolumn, a4size, 11pt]{IEEEtran}
\usepackage{amsmath}
\usepackage{amsthm}
\usepackage{amssymb}
\usepackage{amsfonts}
\usepackage{graphicx}
\usepackage{epsfig}
\usepackage{epstopdf}
\usepackage{subfigure}
\usepackage{psfrag}
\usepackage{cite}
\usepackage{latexsym}
\usepackage{url}
\usepackage{color}
\usepackage{wasysym}
\usepackage[font=footnotesize,skip=1pt]{caption}
\newcommand{\real}{\text{Re}}

\title{Multiuser Charging Control in Wireless Power Transfer via Magnetic Resonant Coupling}
\author{Mohammad R. Vedady Moghadam and Rui Zhang 
\thanks{This paper will be presented in part at IEEE International Conference on Acoustics, Speech, and Signal Processing (ICASSP), Brisbane,
Australia, April 19-24, 2015.}
\thanks{M. R. Vedady Moghadam is with the Department of Electrical and Computer Engineering, National University of Singapore (e-mail: vedady.m@u.nus.edu).}
\thanks{R. Zhang is with the Department of Electrical and Computer Engineering, National University of Singapore (e-mail: elezhang@nus.edu.sg). He is also with the Institute for Infocomm Research, A*STAR, Singapore.}
}
\begin{document}
\maketitle
\thispagestyle{empty}
\begin{abstract}
Magnetic resonant coupling (MRC) is a practically appealing method for realizing the  near-field  wireless power transfer (WPT). 
The MRC-WPT system with a single pair of transmitter and receiver has been extensively studied in the literature, while there is limited work on  the general  setup  with  multiple  transmitters and/or receivers. 
In this paper, we consider a point-to-multipoint MRC-WPT system with  one transmitter sending  power wirelessly to a set of distributed receivers simultaneously. 
We derive the power delivered to  the load of each receiver in  closed-form expression, and reveal a ``near-far'' fairness issue in multiuser power transmission due to users'  distance-dependent mutual inductances with the transmitter.  
We  also show that by  designing  the receivers' load resistances,   the near-far issue can be optimally solved. 
Specifically,  we propose a centralized algorithm to jointly optimize the load  resistances  to minimize the power drawn from the energy source at the transmitter under  given power requirements  for the loads.  
We also devise a distributed  algorithm for the receivers to adjust their load resistances iteratively, for ease of practical implementation. 
\end{abstract}
\newtheorem{definition}{\underline{Definition}}[section]
\newtheorem{fact}{Fact}
\newtheorem{assumption}{Assumption}
\newtheorem{theorem}{\underline{Theorem}}[section]
\newtheorem{lemma}{\underline{Lemma}}[section]
\newtheorem{property}{\underline{Property}} 
\newtheorem{corollary}{\underline{Corollary}}[section]
\newtheorem{proposition}{\underline{Proposition}}
\newtheorem{example}{\underline{Example}}[section]
\newtheorem{remark}{\underline{Remark}}[section]
\newtheorem{algorithm}{\underline{Algorithm}}[section]
\newcommand{\mv}[1]{\mbox{\boldmath{$ #1 $}}}
\section{Introduction}
\label{sec:introduction}
Inductive coupling  \cite{Murakami, Jang, Brusamarello} is a conventional method to realize the near-field wireless power transfer (WPT) for short-range applications up to a couple of centimeters. 
Recently,  magnetic resonant coupling (MRC) \cite{Kurs, Shin, Chen} has drawn significant interests for implementing the  near-field WPT due to its high  power transfer efficiency for applications requiring longer distances, say,  tens of centimeters to several  meters.  
The transmitter and the receiver  in an MRC-WPT system are designed to have the same natural frequency as the system's operating frequency,  thereby greatly reducing the total  reactive power consumption in the system and achieving high power transfer efficiency over long distances.   

The MRC-WPT system with a single pair of transmitter and receiver has been extensively studied in the literature for e.g.   maximizing  the   end-to-end power transfer efficiency or   the power   delivered to the receiver  with a given input power constraint \cite{Cannon,Jonah, YZhang1,YZhang2}. 
However, there is limited work  on analyzing  the MRC-WPT system under the general setup   with multiple  transmitters and/or receivers. 
The system with  two transmitters and a single receiver or a single transmitter and two receivers has been studied in \cite{Yoon,K-Lee, Ahn, Garnica, Johari}, while their analytical  results cannot be applied for a system with more than two transmitters/receivers. 
Furthermore, to our best knowledge,  there has been no work on rigorously establishing a mathematical framework   to jointly design   parameters in the multi-transmitter/receiver MRC-WPT system for its performance optimization.

In this paper, as shown in Fig. \ref{fig:ElecCirtuit}, we consider a point-to-multipoint MRC-WPT  system, where one transmitter connected to a stable energy source sends wireless power simultaneously to a set of distributed receivers, each of which is connected to a given load.
We extend the results in \cite{Yoon,K-Lee, Ahn, Garnica, Johari} to derive  closed-form expressions of the  transmit  power drawn from the energy source and the power  delivered to each load, in terms of  various parameters in the system. 
Our results reveal  a  near-far fairness issue in the case of multiuser wireless power transmission, similar to its counterpart  in wireless communication. 
Particularly,  a receiver that is far away from the transmitter  and thus has a small  mutual inductance with the transmitter generally receives lower power as compared to a receiver that is  close to the transmitter.
We then show  that  the near-far issue can be optimally solved  by jointly designing the receivers' load resistances to control their received power levels, in contrast to  the method of adjusting the transmit beamforming weights to control the received power  in the far-field microwave transmission  based WPT \cite{ZHANG,Xu}. 

Specifically, we  first study the centralized optimization problem, where  a central controller at the transmitter which has the full knowledge of all receivers, including their circuit parameters and load requirements, jointly designs the adjustable load resistances   to minimize  the total power consumed at the transmitter subject to the given  minimum harvested power requirement of each load.  
Although the formulated problem is  non-convex, we  develop an efficient algorithm to solve  it optimally. 
Then, for ease of practical implementation, we consider the scenario without any central controller and devise a distributed algorithm for  adjusting the  load  resistances by individual receivers in an iterative manner.  
In the distributed algorithm, each receiver  sets its load  resistance independently based on its local information and a  one-bit feedback shared by each of the other receivers, where the feedback of each receiver indicates whether the harvested power   of its load  exceeds the required level or not.  
Finally, through simulation results,  it is shown that   the distributed algorithm can achieve close-to-optimal performance as compared to the solution of the 
centralized optimization.  
\section{System model} \label{sec:intro} 
\begin{figure} [t]
\centering
\includegraphics[width=10.5cm]{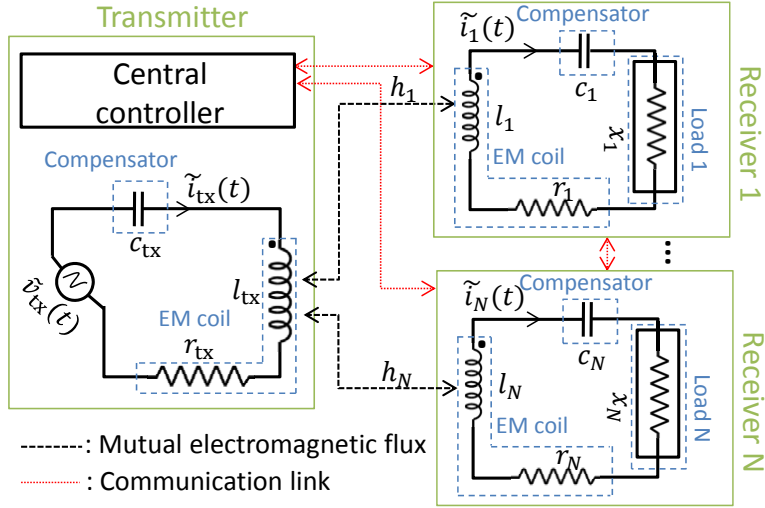}
\caption{A point-to-multipoint MRC-WPT system with communication and control.}
\label{fig:ElecCirtuit} 
\end{figure}

We consider an  MRC-WPT system with  one transmitter and $N$ receivers, indexed by $n$, $ n \in {\cal N}=\{1,\ldots,N\}$,  as shown in Fig. \ref{fig:ElecCirtuit}.  The transmitter and receivers are equipped with  electromagnetic (EM) coils for wireless power transfer.  
An embedded   communication system is  also assumed to enable information sharing among the transmitter and/or  receivers. 
The transmitter is connected  to a stable energy source supplying sinusoidal  voltage  over time given by  $\tilde{v}_{\text{tx}}(t)=\real\{ {v}_{\text{tx}} e^{j w t} \}$,  with   ${v}_{\text{tx}}$ denoting a complex voltage which is assumed to be constant,   and $w>0$ denoting the operating angular frequency of the system.  
Each receiver $n$ is also connected to a given load (e.g. a battery charger),  named   load $n$,  with resistance  $x_{n} > 0$. 
It is  assumed  that the transmitter and each receiver $n$ are compensated  by series capacitors with capacities  $c_{\text{tx}} > 0$ and $c_{n} > 0$, respectively. 
Let $\tilde{i}_{\text{tx}}(t)=\real\{i_{\text{tx}}e^{j w t}\}$,  with complex  ${i}_{\text{tx}} $, denote the steady state current  flowing through the transmitter. 
This current produces a time-varying magnetic flux in the transmitter's EM  coil, which passes through the receivers' EM  coils and  induces time-varying currents in them. 
We  thus denote  $\tilde{i}_{n}(t)=\real\{i_{n}e^{j w t}\}$,  with complex ${i}_{n} $, as the steady state current at receiver $n$.

We denote  $r_{\text{tx}}>0$ ($r_{n}>0$)  and $l_{\text{tx}}>0$ ($l_{n}>0$)  as the internal resistance and the self-inductance of the EM coil of the transmitter (receiver $n$), respectively. 
We  also denote  the mutual inductance between  EM  coils of the transmitter  and each receiver $n$ by $h_{n} > 0$, with $h_{n}  \le \sqrt{l_nl_{\text{tx}}}$,  where its  actual value depends on  the physical characteristics of the two EM  coils,  their locations, alignment or misalignment of their oriented axes with respect to each other,  the environment magnetic permeability, etc.  For example, the mutual inductance  of two coaxial circular loops  that lie in the parallel planes with separating distance of $d$ meter is approximately proportional to $d^{-3}$ \cite{Cheng}.  
Moreover, since the receivers usually employ smaller EM coils than  that of the transmitter due to size limitations and they are also physically  separated,  we can safely ignore  the mutual inductance  between any pair  of them. The equivalent electric circuit model of the considered MRC-WPT system is shown in Fig. \ref{fig:ElecCirtuit},
in which the natural angular frequencies of the transmitter and  each receiver $n$ are given by  $w_{\text{tx}}={1}/{\sqrt{l_{\text{tx}} c_{\text{tx}}}}$ and $w_{n}={1}/{\sqrt{l_{n} c_{n}}}$, respectively. 
We  set  $
c_{\text{tx}}=l_{\text{tx}}^{-1} w^{-2}$ and 
$c_{n}=l_{n}^{-1} w^{-2}$, $\forall n \in \cal N$, to ensure that the transmitter and all receivers have the same natural  frequency as   the system's  operating frequency $w$, named \textit{resonant angular frequency}, i.e.,  $w_{\text{tx}}=w_{1}=\ldots=w_{N}=w$.  

We assume that  the transmitter and all receivers are at fixed positions and the physical characteristics of their EM coils are known; thus,  $h_{n}$, $\forall n\in \cal N$, are modeled as given constants. 
We treat the receivers' load  resistances $x_{n}$, $\forall n \in \cal N$, as  design parameters, which can be adjusted in real-time \cite{Garnica} to control the performance of the MRC-WPT system based on the information shared among different nodes in the system via  wireless communication.  
\section{Performance Analysis}  \label{sec:Performance}
In this section, we first present  our analytical results. 
A numerical example is then provided  to draw useful insights from the  analysis.
\subsection{Analytical Results}
Define $\mv{v}=[v_{\text{tx}},{\bf 0}_{1\times N}]^T$ and $\mv{i}=[i_{\text{tx}},{i}_{1}, \ldots, {i}_{N}]$, where $\mv{v}$ is the  voltage vector  and $\mv{i}$ is the current vector that can be obtained as a function of $\mv{v}$. 
Let $\mv{R}=\text{Diag}( r_{1},\ldots, r_{N})$, $\mv{X}=\text{Diag}( x_{1},\ldots, x_{N})$, and $\mv{h}=[h_1,\ldots,h_N]^T$.
By applying Kirchhoff's circuit laws to the electric circuit model  in Fig. \ref{fig:ElecCirtuit}, we obtain 
\begin{align} \label{eq:V=ZI}
\mv{i}=
\left[ \begin{array}{cc}
r_{\text{tx}}& -j w \mv{h}^{T} \\
-j w \mv{h} & \mv{R}+\mv{X} \end{array} \right]^{-1} \hspace{-1.5mm}
\mv{v}=\mv{A}^{-1} \mv{v},
\end{align}
where $\mv{A} \in \mathbb{C}^{(N+1) \times (N+1)}$ is called the  \textit{impedance matrix}.    The determinant of
$\mv{A}$    is  given by
\begin{align} \label{eq:Determinent}
\text{det}(\mv{A})=\big(r_{\text{tx}}+w^2\sum_{k=1}^{N}h_{k}^2(r_{k}+x_{k})^{-1} \big)\big(\prod_{k=1}^{N} r_{k}+x_{k} \big),
\end{align}
where it can be easily verified that $\text{det}(\mv{A})>0$. 
Then, we define $\mv{B}=\mv{A}^{-1}$, which  is called  the  \textit{admittance matrix}. Let $\mv{B}(b,l)$ denote the element in  row $b$  and  column $l$ of  $\mv{B}$.   We   simplify (\ref{eq:V=ZI}) as   
\begin{align} \label{eq:Simplified_V=ZI}
\mv{i}=\left[\mv{B}(1,1),\ldots,\mv{B}(N+1,1)\right]^{T}v_{\text{tx}}.
\end{align}
It  can also be shown that 
$\mv{B}(b,1)$, $b\in \{1,\ldots,N+1\}$, is given by
\begin{equation} \label{eq:b_mn}
\mv{B}(b,1)\hspace{-0.8mm}=\hspace{-0.8mm} \left\{
\begin{array}{rl}
\hspace{-2mm}~\dfrac{1}{r_{\text{tx}}+ w^2 \sum_{k=1}^{N} h_{k}^2(r_{k}+x_{k})^{-1} } & \text{if}~~b=1,\\
\hspace{-2mm}j\dfrac{w h_{b-1}(r_{b-1}+x_{b-1})^{-1} }{r_{\text{tx}}+ w^2 \sum_{k=1}^{N} h_{k}^2(r_{k}+x_{k})^{-1} } & \text{otherwise}.
\end{array} \right.
\end{equation}
By substituting (\ref{eq:b_mn}) into (\ref{eq:Simplified_V=ZI}), it  follows that
\begin{align} 
i_{\text{tx}}&=~\dfrac{1}{r_{\text{tx}}+ w^2 \sum_{k=1}^{N} h_{k}^2(r_{k}+x_{k})^{-1} }v_{\text{tx}}, \label{eq:IT1m} \\
i_{n}&=\hspace{-0.2mm} j \hspace{-0.2mm}\dfrac{w h_n(r_{n}+x_{n})^{-1} }{r_{\text{tx}}+ w^2 \sum_{k=1}^{N} h_{k}^2(r_{k}+x_{k})^{-1} } v_{\text{tx}}, ~ \forall n \in \cal N. \label{eq:ILm}
\end{align}
The power drawn from the energy source, denoted by $p_{\text{tx}}$, and that delivered to each load $n$,  denoted by $p_{n}$, are then obtained  as
\begin{align} \label{eq:PT}
\hspace{-2.5mm}p_{\text{tx}}&= \hspace{-.5mm}\dfrac{1}{2}\real \{v_{\text{tx}} i_{\text{tx}}^{*}\}
\hspace{-1.1mm}=\dfrac{|v_{\text{tx}}|^2}{2} \dfrac{1}{\hspace{3.3mm}r_{\text{tx}}+ w^2 \sum_{k=1}^{N} h_{k}^2(r_{k}+x_{k})^{-1} \hspace{1mm}}, \\
\hspace{-2.5mm}p_{n}&=\hspace{1.15mm}\dfrac{1}{2} x_{n} |i_{n}|^2 \hspace{1.15mm} \label{eq:PL} \hspace{-0.5mm}=\dfrac{|v_{\text{tx}}|^2}{2} \dfrac{w^2 h_n^2 x_{n} (r_{n}+x_{n})^{-2}}{(r_{\text{tx}}+ w^2 \sum_{k=1}^{N} h_{k}^2(r_{k}+x_{k})^{-1} )^2},
\end{align}
where $i_{\text{tx}}^{*}$ is the conjugate of $i_{\text{tx}}$. From (\ref{eq:PL}), it follows that the power delivered to each load $n$   increases with the mutual inductance between EM coils of its receiver  and  the transmitter, i.e., $h_n$. 
This can potentially cause a near-far fairness issue  since a receiver  that is far away from the transmitter in general has a small mutual inductance with the transmitter; thus, its received power is lower than a receiver that is close to the transmitter (with a larger mutual inductance). 
We accordingly define  $p_{\text{sum}}=\sum_{n=1}^{N}p_{n}$  as the sum (aggregate) power delivered to all loads, where we always have  $p_{\text{sum}}< p_{\text{tx}}$.  

In the following, we study impacts of changing the load resistance of one particular receiver $n$, i.e., $x_n$,  on  the transmitter power  $p_{\text{tx}}$, its received power $p_n$ and that delivered to each of the other loads $m \in \cal N$, $m\neq n$, i.e., $p_m$,  as well as the sum power delivered to all loads $p_{\text{sum}}$,   assuming  that all other load   resistances are   fixed. 
\begin{property} \label{Prop:3}
$p_{\textnormal{tx}}$   strictly increases over $x_n>0$.
\end{property}

This result  can be explained as follows.   From  (\ref{eq:IT1m}), it is  observed that the transmitter current $i_{\text{tx}}$  strictly increases over  $x_n>0$. Hence, due to the fact that the energy source voltage  $v_{\text{tx}}$  is  fixed, it follows that $p_{\text{tx}}$ given  in (\ref{eq:PT})  strictly increases over $x_n>0$. 

\begin{property} \label{Prop:4}
$p_m$,  $\forall m \neq n$,  strictly increases  over  $x_n >0$. However,  $p_n$  first increases  over $0<x_n< \dot{x}_n$, and then decreases over $x_n > \dot{x}_n$, where  
\begin{align} \label{eq:X_Star}
\dot{x}_n=\big({r_n(r_{\textnormal{tx}}+\phi_n)+w^2 h_n^2}\big)/\big({r_{\textnormal{tx}}+\phi_n}\big),
\end{align}
with  
$\phi_n=w^2 \sum_{k\neq n} h_k^2(r_k+x_k)^{-1}$.
\end{property}
 
The above result can be justified as follows.   From  (\ref{eq:ILm}), it follows that  for each receiver $m$, $m \neq n$, its current $i_m$ strictly increases over  $x_n>0$.  This is because $i_{\text{tx}}$ increases with $x_n$, and as a result, $i_m$ increases due to the mutual coupling between EM coils of receiver $m$ and the transmitter.   
Hence,  the received power  $p_m$ defined in (\ref{eq:PL}) also strictly increases over $x_n>0$. On the other hand, it follows from (\ref{eq:ILm}) that for receiver $n$,  its current $i_n$  strictly decreases  over $x_n>0$. Moreover, from (\ref{eq:PL}), it follows that the decrement in $|i_{n}|^2$ is smaller than the increment of $x_n$ when $0< x_n <\dot{x}_n$; thus, $p_n$ increases with $x_n$ in this region, while the opposite is  true when $x_n > \dot{x}_n $.

\begin{property} \label{Prop:5}
If $r_{\textnormal{tx}}+\phi_n-2\varphi_n \le 0$,    $p_\textnormal{sum}$ strictly increases over $x_n>0$, where 
$\varphi_n=w^2\sum_{k\neq n} h_k^2 x_k (r_k+x_k)^{-2}$; 
otherwise,   $p_\textnormal{sum}$  first increases over   $0<x_n<\ddot{x}_n$,  and then decreases over $x_n>\ddot{x}_n$, where  
\begin{align}
\hspace{-2.2mm}\ddot{x}_n\hspace{-1.2mm}=\hspace{-1mm}\big({r_n(r_{\textnormal{tx}}+\phi_n)+w^2h_n^2+2r_n \varphi_n}\big)/\big({r_{\textnormal{tx}}+\phi_n-2\varphi_n}\big).\hspace{-2mm}
\end{align}
\end{property}

This property is a direct consequence of Property 2. 
\subsection{Numerical Example}
 \label{subsec:NumericalExample-Performnce}
\begin{figure}
\centering
\includegraphics[width=12.5cm]{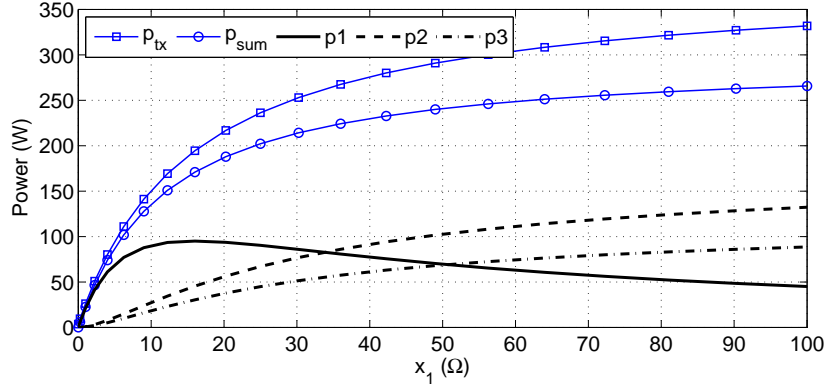}
\caption{Input and output power versus $x_1$.} 
\label{fig:Eff-Pl-versus-rL1} 
\end{figure}

We consider an MRC-WPT system with  $N=3$ receivers, where  $v_{\text{tx}}=25\sqrt{2}$V, $r_{\text{tx}}=0.35 \Omega$, $l_{\text{tx}}=6.35 \mu$H, $r_{n}=0.15\Omega$,  $l_{n}=0.85 \mu$H,  $\forall n \in \cal N$,  $\mv{h}=[2.3, 1.1, 0.9] \mu$H, and $w=2.2 \times 10^6$rad/s. 
In this example, receiver  $1$ is closest to the transmitter and thus it has the largest mutual inductance,  while receiver  $3$ is  farthest.  
For the purpose of exposition,  we fix  $x_{2}=x_{3}=7.5\Omega$.  We plot   $p_{\text{tx}}$,  $p_{n}$, $\forall n \in \cal N$, and   $p_{\text{sum}}$,  versus the resistance of  load $1$, $x_{1}$,  in Fig.  \ref{fig:Eff-Pl-versus-rL1}. 
It is observed that  $p_{\text{tx}}$, $p_2$, $p_3$ and $p_{\text{sum}}$  all increase over $x_1>0$. Note that in this example, the condition $r_{\textnormal{tx}}+\phi_n-2\varphi_n \le 0$  holds in Property 3. 
However,  $p_1$  first increases over $0<x_1<\dot{x}_1=15.8\Omega$, and then  declines  over $x_1>15.8\Omega$. These results are consistent with our above analysis.  
Finally, we point out that changing $x_{1}$ not only affects $p_{1}$, but also  the power delivered to other loads.   For instance,   receiver $1$ can help  receivers $2$ and $3$, which are farther away from the transmitter,  to receive higher power by increasing  $x_1$. This is a useful mechanism that will be utilized to solve the near-far issue.
\section{Centralized Optimization} 
In this section, we  optimize the receivers' load  resistances  $x_{n}$, $\forall n \in \cal N$,   to minimize the power drawn from the energy source at the transmitter subject  to the given load constraints. 
We assume a  central controller at the transmitter, which  has full knowledge of the receivers, including their circuit   parameters and load  requirements,  to implement the proposed centralized optimization. 
\subsection{Problem Formulation}
We  assume that the resistance of each load $n$  can be adjusted over a given range $\underline{x}_n \le x_n \le \overline{x}_n$, where $\underline{x}_n >0$ and $\overline{x}_n \ge \underline{x}_n $ are  lower and upper limits of $x_n$ due to practical considerations. 
We also assume  that the power delivered to each load $n$ should be  higher than a certain power threshold  $\underline{p}_{n}>0$.  
Hence, we formulate the following optimization problem  to minimize the power drawn from the energy source at the transmitter.
\begin{align*} 
\mathrm{(P1)}:  \hspace{-2mm}
&\mathop{\mathtt{min}}_{\{ \underline{x}_n\le x_{n}\le \overline{x}_n\}} 
~\dfrac{|v_\text{tx}|^2}{2}\dfrac{1}{r_{\text{tx}}+ w^2 \sum_{k=1}^{N} h_{k}^2(r_{k}+x_{k})^{-1}}  
\\
\mathtt{s.t.} 
&~\dfrac{|v_\text{tx}|^2}{2}\dfrac{w^2 h_n^2 x_{n} (r_{n}+x_{n})^{-2}}{(r_{\text{tx}}+ w^2 \sum_{k=1}^{N} h_{k}^2(r_{k}+x_{k})^{-1} )^2} \ge \underline{p}_n,~ \forall n \in \cal N. 
\end{align*} 
(P1) is a non-convex optimization problem. 
However,  in the next  we propose an efficient algorithm to solve  (P1) optimally.
\subsection{Proposed Algorithm}
We define an auxiliary variable $z=1/(r_{\text{tx}}+ w^2 \sum_{k=1}^{N} h_{k}^2(r_{k}+x_{k})^{-1})\ge 0$. 
Since $\underline{x}_n\le x_{n}\le \overline{x}_n$, $\forall n \in \cal N$, we   have $\underline{z} \le z\le \overline{z}$, where $
\underline{z}=1/(r_{\text{tx}}+ w^2 \sum_{k=1}^{N} h_{k}^2(r_{k}+\underline{x}_{k})^{-1} )$ and $
\overline{z}=1/(r_{\text{tx}}+ w^2 \sum_{k=1}^{N} h_{k}^2(r_{k}+\overline{x}_{k})^{-1} )$.
Then, we rewrite (P1) as 
\begin{align} 
\mathrm{(P2)}:  
\hspace{+5mm}&\hspace{-5mm}\mathop{\mathtt{min}}_{ \{ \underline{x}_n\le x_{n}\le \overline{x}_n\}, ~\underline{z} \le z\le \overline{z}} 
~{|v_\text{tx}|^2} ~ z /{2} \nonumber \\ 
\mathtt{s.t.} 
&~\dfrac{|v_\text{tx}|^2}{2} z^2w^2 h_n^2 x_{n} (r_{n}+x_{n})^{-2}\ge \underline{p}_n, ~\forall n \in \cal N \label{problem8:const1}  \\
&~r_{\text{tx}}+ w^2 \sum_{k=1}^{N} h_{k}^2(r_{k}+x_{k})^{-1} =z^{-1}. \label{problem8:const2} 
\end{align} 

Although (P2) is still  non-convex, we can solve it in an iterative manner by searching for the smallest  $z$, $\underline{z} \le z\le \overline{z}$,  under which  (P2) is  feasible.
Staring from $z=\underline{z}$,  we test the feasibility  of (P2) given $z$ by considering the  following problem.
\begin{align*}  
\mathrm{(P3)}:  
\mathop{\mathtt{Find}}_{}  
&~ \{ \underline{x}_n\le x_{n}\le \overline{x}_n, ~ 
\mathtt{s.t.}  (\ref{problem8:const1})  \text{ and } (\ref{problem8:const2})\}.
\end{align*} 
If (P3) is feasible,  then we set the optimal objective value of (P2) as $z$,  which can be attained by any  feasible solutions  to  (P3). 
Otherwise,   we set   $z=z+\Delta z$, where $\Delta z>0$ is a small step size. We  repeat the above procedure until  (P3)  becomes feasible or   $z>\overline{z}$. 
The following proposition summarizes the feasibility conditions for  (P3).
\begin{proposition} \label{Prop:4-1}
Given $z$, with $\underline{z} \le z\le \overline{z}$,   (P3) is  feasible  if and only if all conditions listed below hold at the same time: 
\begin{itemize}
\item[C1:]$z \ge 2\sqrt{r_n/\alpha_n}$, $\forall n \in \cal N$, where $\alpha_n=|v_\text{tx}|^2 w^2 h_n^2/(2  \underline{p}_n)$.
\item[C2:]${x}_n^{\text{L}} \le \underline{x}_n\le {x}_n^{\text{U}}$ and/or  ${x}_n^{\text{L}} \le \overline{x}_n\le {x}_n^{\text{U}}$, $\forall n \in \cal N$, where ${x}_n^{L}=( \alpha_n z^2/2-r_n) -z \sqrt{\alpha_n  (\alpha_n z^2/4-r_n)} $ and ${x}_n^{U}=( \alpha_n z^2/2-r_n) +z \sqrt{\alpha_n  (\alpha_n z^2/4-r_n)} $.
\item[C3:]$\Phi=\{  (y_1,\ldots, y_N) ~|~ \underline{y}_n\le y_n \le \overline{y}_n, ~\forall n \in {\cal N}, ~r_{\text{tx}}+w^2 \sum_{k=1}^{k=N} h_k^2 y_k=z^{-1} \} \neq\emptyset$, where  $\underline{y}_n=1/(r_n+\min\{\overline{x},~{x}_n^{U}\})$, and $\overline{y}_n=1/(r_n+\max\{\underline{x},~{x}_n^{L}\})$.
\end{itemize}
\end{proposition}

Given any $(y_1,\ldots,y_n) \in \Phi$, where $\Phi$ is given in C3 of Proposition \ref{Prop:4-1},  the corresponding feasible solution to (P3) is obtained by a change of variable as $x_n=1/y_n - r_n$, $\forall n \in \cal N$. Note that the obtained $(x_1,\ldots,x_N)$ solves (P1) optimally.
To summarize, the  algorithm  to solve (P1) is given  in Table 1, denoted by Algorithm 1.
\begin{table}[t!]
\begin{center} 
\caption{Algorithm  for optimally solving (P1).} \scriptsize{
 \hrule
\textbf{Algorithm 1}
\hrule 
\begin{itemize}
\item[a)] Given $\underline{x}_n>0$ and  $\overline{x}_n> \underline{x}_n$, $\forall n \in \cal N$, compute $
\underline{z}$ and $
\overline{z}$. Initialize $z \gets \underline{z}$,  $\Delta z> 0$, and $Flag \gets 0$. 
\item[b)] {\bf While} $z< \overline{z}$ and $Flag=0$ {\bf do}:
\begin{itemize}
\item[1)] Given $z$, check the conditions listed in Proposition \ref{Prop:4-1}. 
\item[2)] {\bf If} at least one condition does not hold, {\bf then}  set $z=z+\Delta z$.  {\bf Otherwise}, set $Flag=1$.  Choose  any $(y_1,\ldots,y_n) \in \Phi$, where the set $\Phi$ is given in condition C3 of Proposition \ref{Prop:4-1}.  Set $x_n=1/y_n - r_n$, $\forall n \in \cal N$.
\end{itemize}
\item[c)] {\bf If} $Flag=1$, then return $(x_1,\ldots,x_N)$ as the optimal solution to (P1). {\bf Otherwise},   problem (P1) is infeasible.
\end{itemize}
\hrule \label{algorithm:1} }
\end{center}
\end{table}

\section{Distributed  Algorithm} \label{Sec:dist-Alg} 
In this section, we present  a distributed algorithm for  (P1), where it is  suitable for practical implementation when a central controller is not available in the system.   
In this algorithm, each   receiver adjusts its load  resistance independently according to its local information and a  one-bit feedback from each of the other receivers indicating  whether the corresponding load  constraint is  satisfied or not. 
We denote the feedback from each receiver $n$ which is broadcast to all other receivers  as $FB_n \in \{0,1\}$,  where $FB_n=1$ ($FB_n=0$) indicates that its load constraint is (not)  satisfied. 

In Section \ref{sec:Performance}, we show  that   the power delivered to each load  $n$, i.e., $p_n$,  has two properties that can be exploited to adjust  $x_n$. 
First, $p_n$ strictly increases over $x_m>0$, $\forall m \neq n$, which means that other receivers can help boost $p_n$ by increasing their load resistances.
Second, $p_n$ has a single peak at $x_n=\dot{x}_n$, assuming that  other load  resistances are all fixed. 
Thus, over $0<x_n<\dot{x}_n$,  receiver $n$  can increase $p_n$  by increasing  $x_n$; similarly, for $x_n>\dot{x}_n$, it can increase $p_n$ by reducing $x_n$. 
Although  receiver $n$  cannot  compute  $\dot{x}_n$ from (\ref{eq:X_Star}) directly due to its incomplete information on  other receivers, it can  test whether  $0< x_n< \dot{x}_n$, $x_n=\dot{x}_n$, or $x_n>\dot{x}_n$ as follows. 
Let $p_n({x_n^+})$, $p_n({x_n})$, and $p_n(x_n^-)$ be the power received by  load $n$ when its resistance is  set as $x_n+\Delta x$, $x_n$, and $x_n-\Delta x$, respectively, where $\Delta x>0$ is a small step size.  Assuming all the other load resistances are fixed,  receiver $n$ can make the following decision:\\
$\bullet$ If $p_n({x_n^+})>p_n({x_n})$ and $p_n({x_n^-})<p_n({x_n})$, then   $0<x_n<\dot{x}_n$; \\
$\bullet$ If $p_n({x_n^+})<p_n({x_n})$ and $p_n({x_n^-})<p_n({x_n})$, then  $x_n= \dot{x}_n$;\footnote{More precisely, if  $p_n({x_n^+})<p_n({x_n})$ and $p_n({x_n^-})<p_n({x_n})$, then  $\dot{x}_n-\Delta x\le x_n \le \dot{x}_n+\Delta x$.} \\
$\bullet$ If $p_n({x_n^+})<p_n({x_n})$ and $p_n({x_n^-})>p_n({x_n})$, then  $x_n>\dot{x}_n$.

Next, we present the distributed algorithm in detail. 
The algorithm is implemented in an iterative manner, say, starting from receiver 1, where in each iteration, only one receiver $n$ adjusts its load resistance, while all the other receivers just broadcast their individual one-bit feedback $FB_m$, $m\neq n$, at the beginning of each iteration. 
Initialize by randomized  $x_n\in[\underline{x}_n,~\overline{x}_n]$, $\forall n \in \cal N$.  
At each iteration for receiver $n$, if   $p_n < \underline{p}_n$, then it will adjust $x_n$ to increase $p_n$. 
To find the correct direction for the update, it needs to check for its current  $x_n$ whether  $0< x_n< \dot{x}_n$,  $x_n=\dot{x}_n$, or $x_n> \dot{x}_n$ holds, using the method mentioned  in the above. On the other hand,  if $p_n > \underline{p}_n$, receiver $n$ can increase $x_n$  to help increase the power delivered to other loads when there exists any $m\neq n$ such that $FB_m=0$ is received;  or it can decrease $x_n$  to help reduce the transmitter power when $FB_m=1$, $\forall m \neq n$. 
In summary, we design the following protocol (with five cases) for  receiver $n$ to update $x_n$.  \\
C1: \hspace{.5mm}If \hspace{.2mm}$p_n\hspace{.55mm} <\hspace{.5mm}\underline{p}_n$ and $0< x_n< \dot{x}_n$, set  $x_n \gets \min\{\overline{x}_n, x_n+ \Delta x \}$. \\
C2: \hspace{.5mm}If \hspace{.2mm}$p_n\hspace{.55mm} <\hspace{.5mm}\underline{p}_n$  and $x_n> \dot{x}_n$, set  $x_n \gets \max \{\underline{x}_n, x_n- \Delta x\}$. \\
C3: If $p_n> \underline{p}_n$, $x_n \neq \dot{x}_n$, and  $\exists m \neq n$, $FB_m=0$, set $x_n \gets \min \{\overline{x}_n, x_n+\Delta x\}$.  \\
C4: If $p_n> \underline{p}_n$, $x_n \neq \dot{x}_n$, and   $FB_m=1$, $\forall m \neq n$, set $x_n \gets \max \{\underline{x}_n, x_n-\Delta x\}$.  \\
C5: Otherwise, no update occurs. 

In addition, we assume that there is a maximum number of iterations, denoted by  $K_{\max}>1$, after which the algorithm will terminate, regardless of whether it converges to a stable point  $(x_1,\ldots,x_N)$ or not.  However, when the algorithm converges/terminates,  the power constraints  given in (\ref{problem8:const1}) may or may not hold for all loads, depending on the initial values of $x_n$'s. 
If constraint (\ref{problem8:const1}) holds for all loads, then the obtained   $(x_1,\ldots,x_N)$  is a suboptimal  solution to (P1); otherwise, it is infeasible for (P1). 
The  distributed algorithm  is summarized in Table 2, as Algorithm 2. \hspace{-3mm}

\begin{table}[t!]
\begin{center} 
\caption{Distributed algorithm for  (P1).} \scriptsize{
 \hrule
\textbf{Algorithm 2}
\hrule 
\begin{itemize}
\item[a)]  Initialize $Itr=1$ and $K_{\max}\ge1$. Each receiver $n$ randomly chooses $x_n\in[\underline{x}_n,~\overline{x}_n]$.
\item[b)]  {\bf Repeat} from receiver $n=1$ to $n=N$:
\begin{itemize}
\item[--] Receiver $n$  receives $FB_m$ from all other receivers $m\neq n$ and  updates its load resistance $x_n$ according cases C1--C5. 
\item[--] {\bf If} \hspace{1mm}$Itr=K_{\max}$, {\bf then} quit the loop and the algorithm terminates. 
\item[--] Set $Itr=Itr+1$. 
\end{itemize}
\end{itemize}
\hrule \label{algorithm:1} }
\end{center} 
\end{table}  
\section{Simulation Results} \label{sec:Simulation}
\begin{figure}
 \centering
 \includegraphics[width=12.5cm]{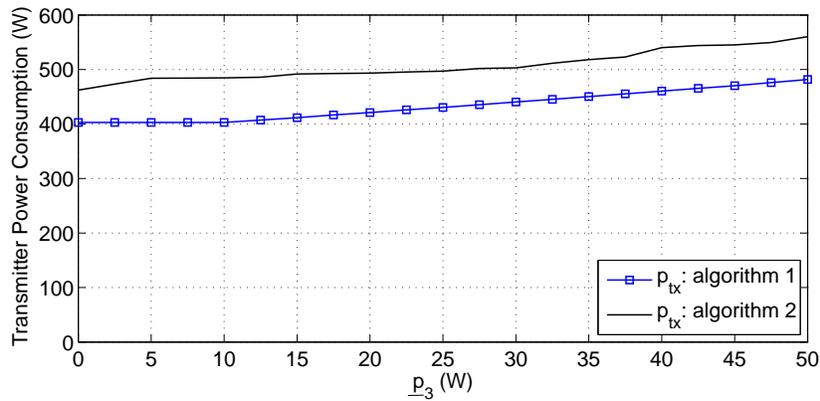} 
 \caption{Performance comparison of Algorithms 1 and 2.}  
 \label{fig:Simul-alg}
 \end{figure}

We consider the  same system setup as that in Section  \ref{subsec:NumericalExample-Performnce}. We set $\underline{x}_n=0.01\Omega$ and $\overline{x}_n=100\Omega$, $\forall n\in \cal N$. 
We also set $\underline{p}_1=250$W, $\underline{p}_2=50$W, and $\underline{p}_3$  varying  as $0 <  \underline{p}_3 \le 50$W.  Note that (P1) is feasible under the above setting. 
For Algorithm 1, we use $\Delta z=10^{-3}$. For Algorithm 2, we use $\Delta x=10^{-3}$ and $K_{\max}=10^5$, which is sufficiently large such that the algorithm converges to a stable point, while  there is no guarantee that the power constraints given in  (\ref{problem8:const1}) hold for all loads at this point. 
Therefore, to evaluate the performance of Algorithm 2, we averaged its result  over 200  randomly generated initial points for each of which the algorithm converged to a feasible solution to (P1).
In Fig. \ref{fig:Simul-alg}, we plot $p_{\text{tx}}$ versus $p_3$.
It is observed that   $p_{\text{tx}}$  obtained by  Algorithm $1$ is lower  than that by  Algorithm $2$, while the gap is quite small,  for all values of $\underline{p}_3$. 
This is  expected since Algorithm $1$ solves (P1)  optimally,  while Algorithm 2 in general only returns  a suboptimal solution. 
\section{Conclusion}
In this paper, we study a point-to-multipoint   MRC-WPT system with distributed  receivers. 
We derive  closed-form expressions for the input and output  power in terms of the system parameters. 
Similar to other multiuser wireless applications such as those in wireless communication and far-field microwave based WPT, a near-far fairness issue is  revealed in our considered system.    
To tackle this problem, we propose a centralized algorithm for jointly optimizing the receivers' load resistances  to minimize the transmitter power  subject to the given load  constraints.  
For ease of practical implementation,   we also devise  a distributed algorithm  for  receivers to iteratively adjust their load resistances based on local information and one-bit feedback  from each of the other receivers.   We show  by simulation   that the  distributed algorithm  performs  sufficiently close to  the centralized algorithm with a finite number of iterations. 
As a concluding remark, MRC-WPT is a  promising research area for which many tools from signal processing and optimization can be  applied to devise new solutions, and we hope that this paper will open up an avenue for future work along this direction.  

\end{document}